\documentclass[12pt]{article}

\usepackage{amsmath}
\usepackage{amssymb}
\usepackage{mathrsfs}
\usepackage{amstext}
\usepackage{graphicx,epsf}
\usepackage{enumerate}
\usepackage{natbib}
\usepackage{color}
\usepackage{multirow}

\newcommand{\G}{\mathcal{G}}

\begin{document}

\begin{center}
\begin{Large}
{\bf Detecting Filamentarity in Climate and Galactic Spatial Point Processes}
\end{Large}

\vspace*{1cm}

\begin{tabular}{lll} 
Aida Gjoka & Robin Henderson$^1$ & Paul Oman \\
Newcastle University, UK & Newcastle University, UK & Northumbria University, UK\\
\end{tabular}
\end{center}

$^1$ Corresponding author: Robin.Henderson\@ ncl.ac.uk

\vspace*{1cm}

{\bf Abstract}

Evidence of excess filamentarity is considered for two spatial point process applications: local minima in whole earth precipitation modelling and locations of cold clumps in the Milky Way. A 
diagnostic test using the number of aligned triads and tetrads is developed.  A Poisson filament process is proposed based on a parent Poisson process with correlated random walk offspring locations.  Filaments are initially identified using an arc search method, with ABC for subsequent inference.  Simulations indicate good performance.  In both applications there is strong evidence of filamentarity.  The method successfully identifies two outlying precipitation data sets.
		
	
	\section{Introduction}

Researchers use very large scale whole-earth models and simulations for the assessment of climate change \citep{kay2015, cast16}.
A common, almost ubiquitous, assumption is that model residuals form a Gaussian random field \citep{cast18, edwards19}.  Recently, topological data analysis methods have been combined with event history techniques to look for departures from the Gaussian assumption beyond marginal properties \citep{garside21}.  These methods are based on the overall topology of a random field. The work in this article arose through our interest in developing similar methods to assess spatial variation and non-stationarity in residual properties in climate data modelling.

\begin{figure}[h!]
\centering{\includegraphics[width = 0.8\textwidth]{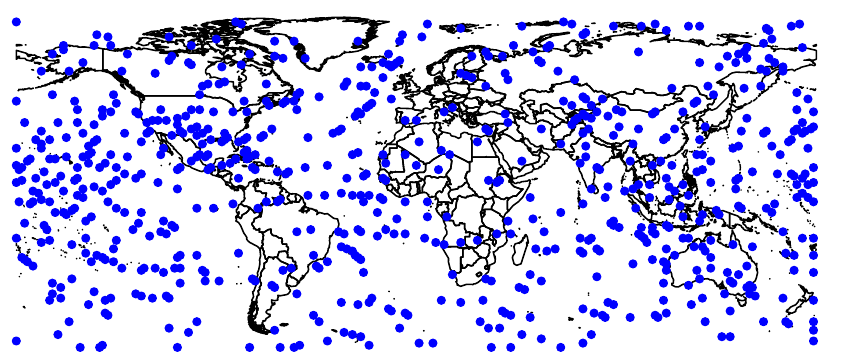}}
  \caption{Precipitation data.}
\label{fig:precip}
\end{figure}

In the context of a random field, the distribution of local maxima and minima forms an important component of a topological data analysis.  Figure \ref{fig:precip}  illustrates, using data from one of 40 climate simulations made available through the Large Ensemble Community Earth System Model (CESM) project, \citep{kay2015}.  The plot shows the global distribution of local minima in annual mean precipitation levels, after allowing for main effects such as latitude, longitude, height, terrain and so on.  In Figure \ref{fig:precip} 
there seems to be a substantial number of {\it filaments}, which are sequences of points that are roughly aligned and not too far apart.  For example the sequence of points that runs from the
northern part of South America across the Atlantic to the coast of Africa. A filament in the pattern of local minima indicates in the climate context a non-isotropic steep sided valley forming a front between two weather systems. 
The same type of pattern occurs for the other ensemble members, though with numbers, lengths and orientations of apparent filaments differing from one repetition to another. The presence of filamentarity in a point pattern is an aspect of non-Gaussian random fields that has been widely studied in cosmology and astrophysics \citep{schmal99, novi06, libes17}. If there is indeed stochastic filamentarity in the spatial distribution of topological features for climate data residuals then the standard Gaussian random field assumption fails.

As pointed out by \cite{bhara2000} however,  the eye is susceptible  to picking out structure when in fact there is none.  We need therefore a method to 
distinguish in plots like  Figure \ref{fig:precip} a genuine excess of filamentarity from an apparent excess. Further, a model that allows us to characterise a stochastic filamentarity process will allow the exploration of similarities and differences between the 40 climate model realisations.  Consistency between ensemble members is an important quality assurance requirement for large simulation models such as CESM \citep{baker15}.

\begin{figure}[h!]
\centering{\includegraphics[width = 0.6\textwidth]{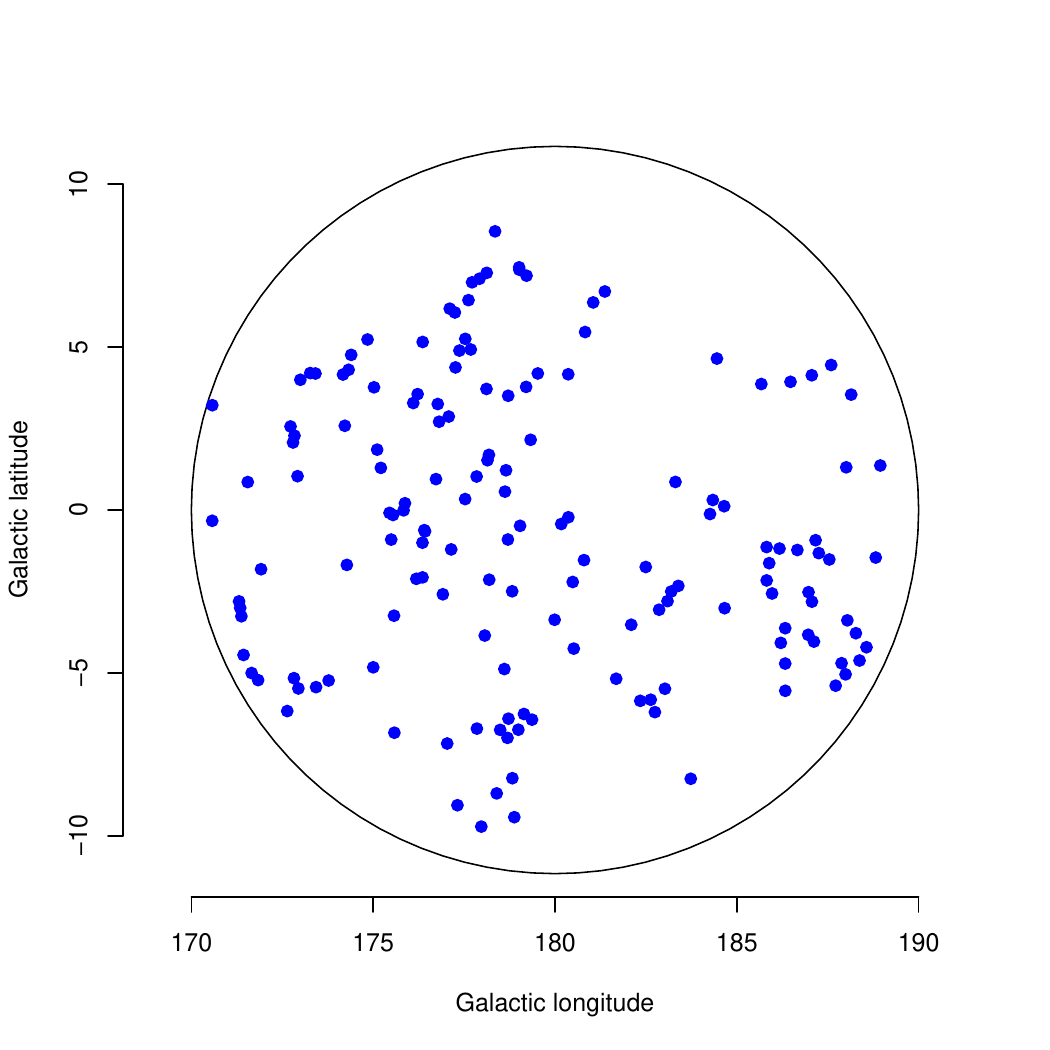}}
  \caption{Cold clump data. }
\label{fig:coldclump}
\end{figure}

The identification of filamentarity within point processes is a problem of longstanding  and continuing interest to astrophysicists, with some 40\% of galaxies aligned along the cosmic web. For examples see \cite{edmunds85}, \cite{darvish17}, \cite{bhara2000}, \cite{xu19} or \cite{kuutma20}, and for a comprehensive review see  \cite{libeskind18}.  Recently  published data  brings the problem closer to home and provides our second application: the possible alignment of cold clumps within our own galaxy. Cold clumps are regions of dense and relatively cold dust that are the precursors to star formation. 
As well as filamentarity in the distribution of galaxies, there is ongoing discussion as to whether stars within a galaxy may also show alignment as a result of filamentarity in the molecular clouds from which they are formed \citep{myers17, lu18}.
The Planck collaboration has now catalogued  cold clumps within the Milky Way \citep{planck16}, which means we can investigate whether there is evidence of filamentarity in their locations and hence provide support or otherwise for the suggestion of star alignment.  Figure \ref{fig:coldclump} shows all 
cold clumps in a disc of radius 10$^\circ$ centred at the galactic origin of the Milky Way, a region which can be considered fairly homogenous. Whether there is a filamentary structure to cold clumps at the intragalactic scale is a question that has not to our knowledge so far been considered statistically.

For both the precipitation data and cold clump applications a first task is to determine whether there are more filaments in the observed patterns than might be expected under the common assumption of a locally stationary Poisson process.  A variety of statistical methods have been developed, often based
on the  intensity of points within specified geometric shapes (rectangles, tubes)  \citep{hall06, stoica07, genovese12, chen15, qiao16}.  The methods work well when the number of points is large and the purpose is to identify filament-shaped regions of high intensity.  Our focus is on individual points in a cloud of modest size and our starting point is the blunt triangle work of \cite{broadbent80} and \cite{kendall80}.  This is developed in 
Section \ref{sec:blunts}.

In Section \ref{sec:pfp} we propose a modelling procedure that can be used for simulation and inference, akin to a Poisson cluster process \citep{cressie15} but with  offspring following a correlated random walk rather than being randomly scattered around their parent location.  The observed data likelihood for this process is intractable but simulation is trivial, which makes approximate Bayesian computation \citep{fearnhead12} an obvious choice of estimation technique.  This is developed in Section \ref{sec:est} after we first describe in Section \ref{sec:arc}  a simple but effective method for the initial connection of points into filaments. Our two applications will be presented in 
Section \ref{sec:apps}.  The precipitation data are defined on a sphere and this will properly be taken into account when we consider these data in detail.  For the most part however we will consider the more generic problem of point patterns in a two dimensional plane, with Euclidean distances.

\section{Blunt triads}
\label{sec:blunts}
\subsection{Methodology}

As a simple initial diagnostic we propose counting the number of triads of points that are approximately aligned, and comparing this with that  expected under a stationary and homogeneous Poisson process over the same two-dimensional region. By approximately aligned we mean that the triangle defined by the three points has largest angle greater than $\pi-\varepsilon$, while the edge lengths on either side of the largest angle are both less than $d_0$.  Here $\varepsilon$ and $d_0$ can be user-defined and will usually be relatively small.  

If edge length is ignored, a triangle with maximum angle  greater than $\pi-\varepsilon$ was called $\varepsilon$-{\it blunt} by \cite{kendall80}.  
Following \cite{broadbent80}, \cite{kendall80} consider properties of the number $N(\varepsilon)$ of $\varepsilon$-blunt triads  amongst $n$ points randomly distributed over a region $K$ of the plane, with emphasis on the mean and variance  of $N(\varepsilon)$ as $\varepsilon$ decreases.  We adapt their methods here to allow the extra restriction on edge length, letting
$N(\varepsilon, d_0)$ be the number of $\varepsilon$-blunt triads that also satisfy the adjacent edge restriction.  

\begin{figure}[h]
\centering{\includegraphics[width = 0.65\textwidth]{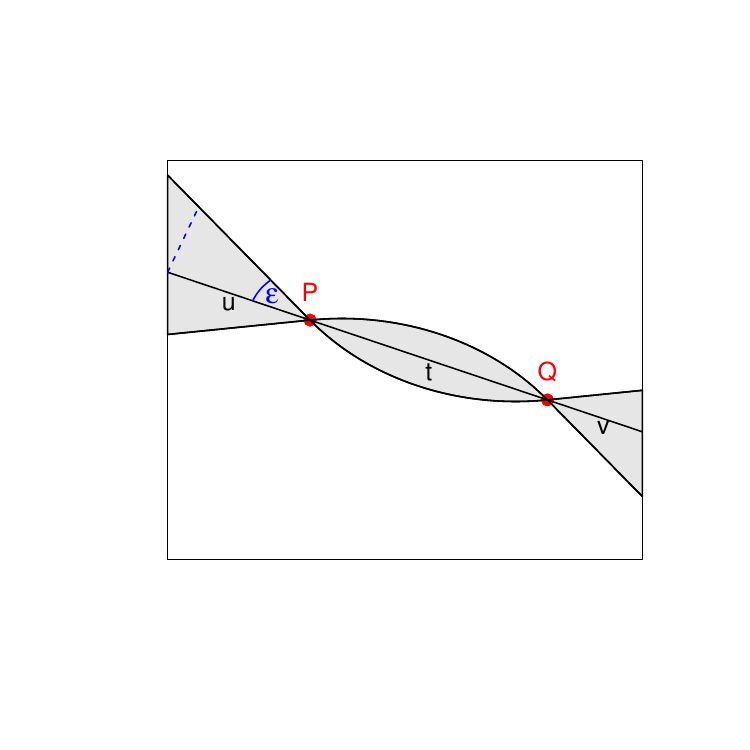}}
  \caption{Blunt triangle region and secants defined by angle threshold $\varepsilon$ and random points $P$ and $Q$}
\label{fig:secants}
\end{figure}

A starting point is to consider two points $P$ and $Q$ in $K$ and determine the probability that a third point $R$ falls in a region that leads to an 
 $(\varepsilon,d_0)$-blunt triad.  Figure \ref{fig:secants} illustrates, for the case of $K$ being a rectangle. Here the shaded region indicates locations of $R$ that would lead to an $\varepsilon$-blunt triad irrespective of edge lengths.  \cite{broadbent80} calls the end regions ``wedges'' and shows that their areas are $u^2 \varepsilon+O(\varepsilon^2)$ and  $v^2 \varepsilon+O(\varepsilon^2)$, while the central ``lens'' region has area
$\frac{1}{3}t^2 \varepsilon+O(\varepsilon^2)$.  Adding the additional edge constraints leads to
\[ \Pr(\mbox{Triangle } PQR \mbox{ is } (\varepsilon,d_0)\mbox{-blunt} \mid P, Q)  =  \frac{1}{|K|} H(P, Q) \varepsilon +O(\varepsilon^2),\]
 where 
\[ H(P,Q) =  \left\{ \begin{array}{ll}
u_{d_0}^2 + \frac{1}{3}t^2 +v_{d_0}^2  & \mbox{if } t < d_0,\\
\\
2d_0^2- \dfrac{t^2}{3} - \dfrac{4d_0^3}{3t}  &  \mbox{if } d_0 \leq t \leq 2d_0,\\
\\
0 &  \mbox{if }  t > 2d_0,\\
\end{array} \right.
\]
and $u_{d_0}= \mathrm{min}(u,d_0)$, $v_{d_0} = \mathrm{min}(v,d_0)$.
The derivation is outlined in Appendix A.

Adapting~\cite{kendall80}, let $p(\varepsilon, d_0)$ be the probability that three points independently uniformly distributed over $K$ form an  $(\varepsilon,d_0)$-blunt triad.  Define  $\alpha(K) = \lim p(\varepsilon, d_0)/\varepsilon$ as $\varepsilon \downarrow 0$.  Then 
\begin{equation}
 \alpha(K) = \frac{1}{|K|}\mathbb{E}_P\mathbb{E}_Q H(P,Q).
\label{eqn:lambda}
\end{equation}
It is straightforward to calculate $\mathbb{E}_P\mathbb{E}_Q H(P,Q)$ by numerical integration, or by Monte Carlo methods given that simulation of $P$ and $Q$ is trivial.  In our numerical work we used Monte Carlo with 100,000 replicated $(P,Q)$, which was almost instant.

If we now assume $n$ points randomly distributed over $K$, the expected number of $(\varepsilon,d_0)$-blunt triads is 
\[ \mathbb{E}[N(\varepsilon, d_0)] = {n \choose 3} \alpha(K)\varepsilon+O(\varepsilon^2). \]
The variance of $N(\varepsilon, d_0)$ also follows from \cite{kendall80} with our new $H(P,Q)$ in place. The derivation is given for completeness in Appendix A and the final result is 
\begin{eqnarray}
\mbox{Var}[N(\varepsilon, d_0)] & = & \mathbb{E}[N(\varepsilon, d_0)](1-\alpha \varepsilon ) 
  + 3 {n \choose 3}{n-3 \choose 2}(\beta-\alpha^2)\varepsilon^2 \nonumber\\
 & + &  3 {n \choose 3}{n-3 \choose 1} (\gamma-\alpha^2)\varepsilon^2+O(\varepsilon^3),
\label{eqn:varN}
\end{eqnarray}
where $\alpha=\alpha(K)$ as previously defined and 
\[ \beta = \beta(K) =  \frac{1}{|K|^2}\mathbb{E}_P\{[\mathbb{E}_Q H(P,Q)]^2\},\]
\[ \gamma = \gamma(K) =  \frac{1}{|K|^2}\mathbb{E}_P\mathbb{E}_Q\{ H(P,Q)^2\}.\]

Numerical work indicates that the expected value and variance expressions are accurate.  Table \ref{tab:maxedgesim} illustrates for $n=40$ points in a $s \times 1$ rectangle $K$. The theoretical quantities derived above are compared with empirical values from 1000 replications and results are good.  Note that the theoretical values are easy to calculate for any $n$, but simulation becomes infeasible for large $n$ and large $d_0$ because of the number of triads that have to be assessed.  For example at $n=697$ as for the precipitation data, over 50 million triads would have to be examined for each individual simulation.

 \begin{table}
 \caption{\label{tab:maxedgesim} Comparison of theoretical and simulation expected value and coefficient of variation CV of $N(\varepsilon, d_0)$ at $n=40$.  Here $A=\varepsilon\pi/180$ is $\varepsilon$ measured in minutes.}
 
  \centering 
  
 \begin{tabular}{rllrrrrrr} 
		  & & & \multicolumn{2}{c}{$d_0=\infty$} & \multicolumn{2}{c}{$d_0=0.5$} & \multicolumn{2}{c}{$d_0=0.25$}\\ 
		 $A$ & $s$& & $\mathbb{E}(N)$ & CV$(N)$ & $\mathbb{E}(N)$ & CV$(N)$ & $\mathbb{E}(N)$ & CV$(N)$\\ 
	 
		 10 & 1 &Theory   & 9.58 & 0.33 & 5.12 & 0.47 & 0.65 & 1.26\\ 
		 10 & 1 &Simulation &9.70 & 0.34 & 5.26 & 0.44 & 0.67 & 1.27\\ 
		 10 & 3 &Theory  &15.95 & 0.27 & 0.91 & 1.07 & 0.09 & 3.44\\ 
		 10 & 3 &Simulation  & 16.22 & 0.26 & 0.87 & 1.13 & 0.12 & 2.72\\ 
		 60 & 1 &Theory   & 57.31 & 0.15 & 30.57 & 0.23 & 3.90 & 0.57\\ 
		 60 & 1 &Simulation  &  57.36 & 0.15 & 30.07 & 0.24 & 3.90 & 0.53\\ 
		 60 & 3 &Theory   & 95.95 & 0.15 & 5.44 & 0.48 & 0.51 & 1.44\\ 
		 60 & 3 &Simulation  & 95.03 & 0.14 & 5.70 & 0.43 & 0.53 & 1.38\\ 
\end{tabular} 
 \end{table}

\subsection{Results}
\label{sec:firstres}


Table \ref{tab:firstres} compares the observed numbers of $(\varepsilon, d_0)$-blunt triads in the precipitation and cold clump data with the expected properties should the points form a homogeneous Poisson process. The theoretical properties for the cold clump example are adjusted to allow a circular observational region. 
 In both cases we took $\varepsilon=15\pi/180$. For the precipitation data we took $d_0=10$ and for the cold clump data $d_0=\sqrt{2}$.  In both applications there are many more $(\varepsilon, d_0)$-blunt triads than would be expected under a Poisson process.

We also considered aligned tetrads.  These are ordered sequences of four points $P_1P_2P_3P_4$ with the two interior angles greater than $\pi-\varepsilon$ and with all edge lengths $P_iP_{i+1}$ less than $d_0$.  Given that aligned triads have been identified, it is straightforward to find aligned tetrads: if $P_1P_2P_3$ is an aligned triad then $P_1P_2P_3P_4$ is an aligned tetrad if and only if $P_2P_3P_4$ is itself an aligned triad.  We have no theoretical values for tetrads but once more the observed values are much higher than values seen in simulated Poisson processes.

Thus we conclude that we have evidence of filamentarity in both applications and we can proceed to more thorough modelling and analysis.

\begin{table}
\caption{  \label{tab:firstres} $N(\varepsilon, d_0)$  for  cold clump and precipitation data compared with theoretical values under a homogeneous Poisson process.} 
\centering 
 
\begin{tabular}{llcrrrr}
     & &           & \multicolumn{2}{c}{Theory} & \multicolumn{2}{c}{Simulation}\\
&     & Observed &  $\mathbb{E}(N)$ & CV$(N)$ & $\mathbb{E}(N)$ & CV$(N)$ \\
 
Precipitation &Triads & 812 & 441.1 & 0.10 & 438.2 & 0.08\\
&Tetrads& 428 & - & - &  141.4 & 0.18\\
\\
Cold clumps &Triads & 126 & 63.0 & 0.20 &62.6 & 0.24\\
&Tetrads& 34 & - & - &  17.0 & 0.56\\
\end{tabular} 
 
\end{table}

\section{A Poisson Filament Process}
\label{sec:pfp}

\begin{figure}[h!]
\centering{\includegraphics[width = 0.8\textwidth]{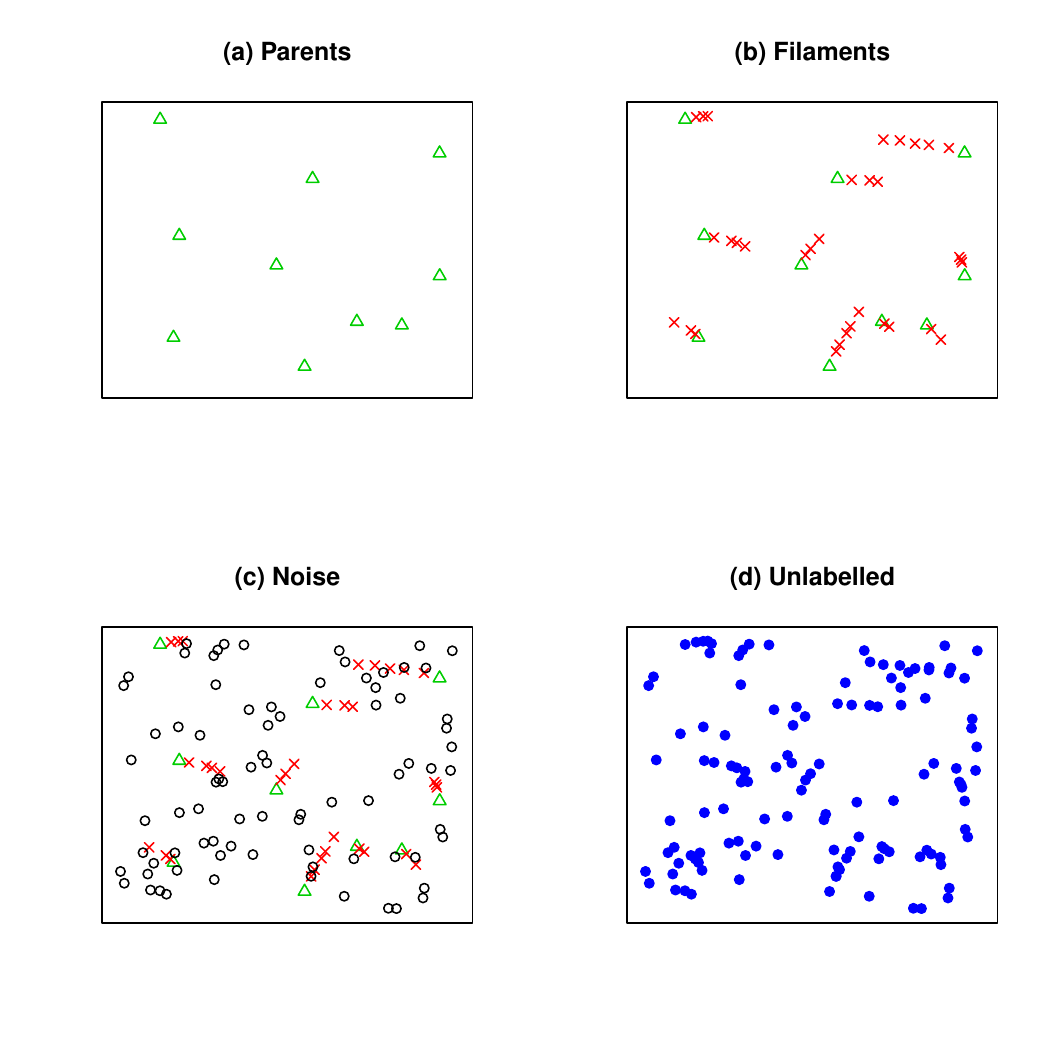}}
  \caption{Poisson filament process.}
\label{fig:pfpex}
\end{figure}

The Poisson cluster process is familiar \citep{cressie15}. In the simplest case parents are realised from a homogeneous Poisson process over a region $K$, each parent generates, independently,  a random number of offspring, whose locations relative to their parent are independently drawn from some distribution. We adapt this to form a Poisson filament process as follows.
\begin{enumerate}
\item Filament parents are generated from a homogeneous Poisson process over  $K$. 
The number of filaments has Poisson distribution with parameter $\lambda_0$. Panel (a) of Figure \ref{fig:pfpex} illustrates. 
\item The number of points in filament $i$ is $N_i=3+M_i$, where the $M_i$ are independent  random variables from a distribution $g_M$ on the non-negative integers. This ensures that filaments have length at least three, including the parent.
\item Starting from the parent location, the points form a correlated random walk, as in Panel (b) of Figure \ref{fig:pfpex}.  The initial direction is drawn from some distribution $g_{0}$, the edge lengths are independent from distribution $g_L$, and the {\it changes} in direction are independent from a third distribution $g_C$. The three distributions  $g_{0}$, $g_L$, and $g_C$ can be parametrised as desired.
\item Noise is added either as  a homogeneous Poisson process with parameter $\lambda_1$ (Panel (c)) or as a specified number of points independently uniformly distributed over $K$.  The latter is useful if the total number of points is to be fixed.

\item The points are not labelled (Panel (d)).
\end{enumerate}

\subsection{Simulation I}
\label{sec:sims1}

We performed a simple simulation experiment to investigate the properties of  the numbers of blunt triads or tetrads as test statistics, aimed at distinguishing a Poisson filament process from either a homogeneous Poisson process or a comparable Poisson cluster process.  The simulations loosely mimic the precipitation data, with all realisations having a fixed total of $N_{tot}=697$ points and $K$ being a $150 \times 360$ rectangle.  The number of points in a filament was uniformly distributed on $\{3,4,\ldots,8\}$, the edge lengths had $U(2,d_0)$ distribution, the initial direction was $U(0,2\pi)$ and the subsequent changes in direction were independent $U(-\varepsilon,\varepsilon)$ variables.  We took $d_0=10$ and $\varepsilon=15\pi/180$ and used the same values in searching for blunt triads and tetrads, as in Section \ref{sec:firstres}.
We generated data with a varying proportion $w$  of points in filaments.  At $w=0$ we have a pure Poisson process and at $w=1$ we have only filaments, with no noise.

Results are shown in Table \ref{tab:sims1}.  We took either a Poisson process or a comparable Poisson cluster process as null hypothesis, with a Poisson filament process as alternative.  Properties under both null and alternative were based on 1000 replications for each value of $w$.  We see that the numbers of blunt triads and tetrads each have excellent power to identify even a small amount of filamentarity compared with a pure Poisson process, but there is  no power in these statistics which would allow us to distinguish a Poisson filament process from a Poisson cluster process. The reason is that short filaments (triads or tetrads) are common by chance in clustered data.  We return to this issue later.

\begin{table}
\caption{\label{tab:sims1}Size and power of test for filamentarity based on blunt triads or tetrads. PP is Poisson process and PCP is Poisson cluster process}
  \centering 
  
\begin{tabular}{lrrrr}
 & \multicolumn{2}{c}{$H_0$: PP} &  \multicolumn{2}{c}{$H_0$: PCP}\\ 
$w$   &  Triads & Tetrads & Triads & Tetrads\\
0     & 0.06  &  0.05 & 0.04 & 0.03\\
0.05  & 0.45  &  0.66 & 0.01 & 0.04\\
0.1   & 0.87  &  0.97 & 0.00 & 0.05\\
0.15  & 0.99  &  1.00 & 0.00 & 0.05\\
0.2   & 1.00  &  1.00 & 0.00 & 0.04\\
0.25  & 1.00  &  1.00 & 0.00 & 0.03\\
\end{tabular} 
\end{table}

\section{Initial filament identification}
\label{sec:arc}



In order to distinguish between a Poisson filament process and a Poisson cluster process, we, therefore, require a different approach that focusses more on detecting linear runs of points. Our new approach, called arc search (AS), looks to capture sequences of points that satisfy the blunt triangle conditions and knit them together to create longer linear filaments. 

First, consider a point $x_i, i= 1, 2, \dots, n $ in a two-dimensional plane, and  locate its nearest neighbour, point $x_j$ say. If the distance between $x_i$ and  $x_j$ is greater than a predetermined threshold  $d_0$, then we move on to the next point  $x_{i+1}$ (or stop if  $i=n$). However, if the distance between the points is not greater than $d_0$, then we join these two points and consider this edge as being the start of a potential filament. To make this algorithm clearer,  consider two such points, renamed $P$ and $Q$ in Figure \ref{fig:secants}. 

To see if our filament can be extended, we will next project a line from $P$ through $Q$ for a distance $d_0$ and look to see whether any points lie near this projected line. Ideally, the next point will lie on the projected line, however we could accept points that are within a short distance of the projection as they could still form a filament as long as the angle formed at $Q$ is at least $\pi - \varepsilon $. To detect all possible points that could extend the potential filament $PQ$, we will create an  \textit{arc of acceptability} by looking at a segment of side $d_0$ which stretches by an angle $\varepsilon$ either side of the projected line from $Q$ with all points in the arc of acceptability no greater than $d_0$ units from $Q$. Examining Figure \ref{fig:secants}, we see that the region under discussion will contain at least part of the line segment $v$ and the shaded region around it. If there are no points in this region, then we cannot extend the filament beyond $Q$ and so we will stop searching in this direction and look to see if we can extend the filament in the opposite direction from $Q$ to $P$ in the shaded region that contains at least part of the line segment $u$. We will apply the same algorithm as above and will continue extending the filament until the arc of acceptability fails to locate any new acceptable points.

If the arc of acceptability identifies one qualifying candidate, then that point joins the filament: we could label this as point $R$. Then $PQR$ forms a blunt triangle and we would now look to extend the filament $PQR$ in the same direction using the search description above but beginning with the edge $QR$.

If the arc of acceptability identifies more than one qualifying candidate, then the point nearest to the projected line is selected to extend the current filament. We only select one point in such a  situation as we want to create a filament with the most linear set of points and reduce unnecessary noise. This is equivalent to selecting the blunt triangle with the largest obtuse angle. Any points ignored at this stage could still be included in other filaments created in other searches.

We start the algorithm at every point in the data set in turn and collate all of the filaments that satisfy the length and angle conditions.  The procedure can yield a long list of filaments of varying lengths, depending on the parameters $\varepsilon$, $d_0$ and $n$, so we next look to rationalise. All filaments consisting of just two points are removed, as are any filaments that are a subset of another filament. Finally, we  join filaments with end points in common to create longer filaments where possible. Following the algorithm produces a set of filaments that successfully identify most linear structures of points that satisfy distance and angle restrictions.

An alternative approach to analysing the spatial distribution of points in a two-dimensional plane is to use the Minimal Spanning Tree (MST). These trees are helpful because they create a unique footprint of the data, subject to mild conditions, and summary measures can be created from the MST to highlight pertinent features of the data. The MST was first used in the study of large scale galaxy structures by \citet{barrow85}, who developed several statistical tests to compare distributions of points in galaxies against a Poisson process assumption. We shall return to one of these tests later and we note that MSTs have also been used in many other studies to analyse arrays of points, including \citet{krzewina96},  \citet{campana13},  \citet{pereyra20} and \citet{naidoo20}. Instructions on how to construct an MST can be found within each of these articles. In \texttt{R},  the \texttt{ComputeMST} function in the \texttt{emstreeR}   package yields the MST. A concern when attenpting to detect linear filaments using MSTs is that the MST can only detect filaments that run along its branches. We will compare the  AS and MST approaches to detecting filaments in the next section.

\subsection{Simulations II}

We generated point patterns using the  scenario of  Section \ref{sec:sims1}, except that five different levels of the mixture parameter $w$ were considered: $w \in \{0.1, 0.3, 0.5, 0.7, 0.9\}$.  The positions of the true filaments were noted and then filaments were estimated using both the AS routine and the MST. The results below are based on batches of $500$ simulations.

\subsubsection{Filament identification}

Given that points are not labelled, the random points added as noise in the data generation  clearly complicate any filament search technique. They could extend existing true filaments, create new filaments or  add points in the middle of the true filaments. Therefore we first examine how often over half of each true filament is allocated to a single estimated filament. For example, for true filaments consisting of three points, the estimated filament would have identified at least two of the points. For filaments consisting of four or five points, we would be registering those estimated filaments that had identified at least three of the points. By setting the identification threshold at 50\%, the estimated filaments give us a sense of where the true filaments are, even if we do not capture every point. Table \ref{tab:sims2} displays the results. We see that the AS method consistently outperforms the MST method in detecting the true filaments.

\begin{table}
	\caption{ \label{tab:sims2}Proportion of estimated filaments capturing  over $50\%$ of true filaments. Values given are the  median and range over 500 replications}
\centering
	\begin{tabular}{lcc }
	 
		$w$   &  AS & MST \\
		0.1    &0.92  &   0.67   \\
		&(0.50-1.00)&(0.17-1.00) \\
		0.3  &  0.92 &   0.67  \\
		& (0.72-1.00)&(0.39-0.91) \\
		0.5  &  0.92  & 0.68 \\
		&(0.81-1.00)&(0.45-0.86) \\
		0.7  &  0.93&  0.69 \\
		& (0.84-1.00) &(0.51-0.85) \\
		0.9   & 0.93 & 0.70 \\
		&(0.86-0.98) &(0.58-0.85) \\
		
	\end{tabular}
\end{table}

\subsubsection{Sensitivity and Specificity}

We can also look at the individual points on the true filaments to see how often they are detected by each method. As a sensitivity measure we consider the proportion of points on true filaments that are also on estimated filaments.  For specificity we determine the proportion of the noise points, namely not on true filaments,  that are also not on estimated filaments.  Table  \ref{tab:sims21} summarises our results.

The AS method has much higher sensitivity than the MST methods. Points that are in true filaments are more likely to be in estimated filaments if the AS method is preferred to MST.  The latter is more conservative, leading to fewer estimated filaments and consequently greater specificity.\\

\begin{table}
\caption{\label{tab:sims21} Sensitivity and specificity. Values given are the medians and range over 500 replications}	
\centering
\begin{tabular}{lcccc}
		& \multicolumn{2}{c}{Sensitivity} &  \multicolumn{2}{c}{Specificity}\\ 
		$w$ &  AS & MST & AS & MST\\
		0.1 & 0.96  &  0.69 & 0.42 & 0.65\\
		    &(0.81-1.00)&(0.42-0.94)&(0.33-0.56)&(0.56-0.74)\\
		0.3 &  0.95  &  0.69 & 0.45 & 0.68\\
		    & (0.88-0.99)&(0.52-0.81)&(0.34-0.57)&(0.59-0.78)\\
		0.5 &  0.96  & 0.69 &0.47 & 0.70\\
		    &(0.9-1.00)&(0.57-0.81)&(0.32-0.62)&(0.60-0.81)\\
		0.7 &  0.96 &  0.70 & 0.49 & 0.72\\
		    & 0.91-0.98) &(0.60-0.80)&(0.35-0.61)&(0.60-0.83)\\
		0.9 & 0.96 & 0.70 & 0.51 & 0.75\\
		    &(0.92-0.99) &(0.62-0.80)&(0.29-0.73)&(0.56-0.92)\\
	\end{tabular}
\end{table}

\subsubsection{Using AS Filaments as a filamentarity diagnostic}
 
We now consider discrimination between linear clusters of points and cloud-like clusters of points. As seen previously, there is no power in blunt triangles or blunt tetrads for this problem. We will use a linearity measure introduced by \citet{barrow85}, that compares the sum of the edge lengths on a filament to its end-to-end separation. Values will approach one for near-linear filaments but will be higher for non-linear filaments. We will have a linearity measure for each filament identified for a set of data and a convenient metric for our simulations is the median linearity measure over the identified filaments.

Simulations again followed the scenario of Section \ref{sec:sims1}, this time with batches of 100 replicates for each considered parameter combination. First, we generated completely random sets of points in this region to mimic a purely Poisson process and then fitted arc filaments to each data set using values $d_0=10$ and $\varepsilon = 15 \pi /180$. For each set of filaments, we calculated the median linearity score and used these scores to determine one-sided critical values from the probability distribution of median linearity scores for filaments fitted to a random Poisson process. 

\begin{figure}[h!]
	\centering{\includegraphics[width = 0.55\textwidth]{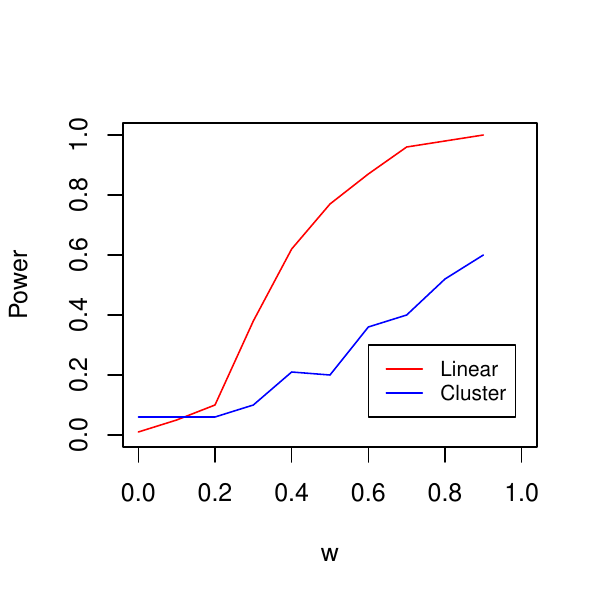}}
	\caption{Power plot of linearity measure of filamentarity.}
	\label{fig:linearity}
\end{figure}

We then introduced some filamentary structure to the data sets using the method detailed above. We looked at different levels of true filament level measured through parameter $w$, taking values from $0$ to $0.9$ in increments of $0.1$,  and generated either a Poisson filament process or a comparable Poisson cluster process. For each simulation, we compared the median linearity measure with the critical values and by accumulating these comparisons, we were able to estimate the power of correctly identifying non-random structure in the simulated data. Figure \ref{fig:linearity} illustrates the results.

We see that the median linearity measure has high power to detect structure in a point process for even small levels of filamentarity, whereas power is low when the alternative is a Poisson cluster process. A statistically significant result would point toward filamentarity rather than clustering, giving more discriminatory ability than the counts of blunt triads and tetrads explored in Section 3.1.
 We can, therefore, use this measure to compare the effectiveness of arc filaments in detecting the presence of linear filaments in data sets.

\section{Estimation}
\label{sec:est}

We turn now to parameter estimation for a Poisson filament process. Let $\theta$ be the parameter vector of interest. Recalling that filaments have size $3+M$ with $M \sim g_M$, we take $g_M$ to be the Poisson distribution with mean $\mu$ from now.  We assume that the main interest is in three parameters: the filament number parameter $\lambda_0$, the noise parameter $\lambda_1$, and the filament size parameter $\mu$.

The observed data likelihood is not tractable but simulation is very straightforward, meaning an approximate Bayesian computation (ABC) method is attractive. We give only a summary of the method here: 
see \cite{fearnhead12} for an excellent description of ABC techniques in general. We suggest the following algorithm.

\begin{enumerate}
\item[1.] Perform an arc search on the observed data $y_{{\tt obs}}$ to obtain initial estimated filaments, as in Section \ref{sec:arc}, and with the requirement that no point can be in more than one filament.
\item[2.] Prune any excess of short filaments, i.e. of length three.  This can be achieved by fitting a zero-inflated Poisson distribution to the empirical filament size distribution (minus 3), and randomly deleting filaments of length three until there is no statistically significant excess. 
\item[3.] Map the points and estimated filaments to a feature vector $s_{{\tt obs}}$.
\item[4.] Generate a parameter vector $\theta$, either from an appropriate prior or as a Markov chain.   
\item[5.] Simulate data $y_{{\tt sim}}$ from a Poisson filament process with parameter $\theta$.
\item[6.] Repeat steps 1-3 on $y_{{\tt sim}}$ to obtain a feature vector $s_{{\tt sim}}$.
\item[7.] Determine a distance $d(y_{{\tt sim}},y_{{\tt obs}})$ between $s_{{\tt sim}}$ and $s_{{\tt obs}}$.
\item[8.] If $d(y_{{\tt sim}},y_{{\tt obs}})$ is below a specified threshold accept $\theta$, otherwise reject it.
\item[9.] Iterate steps 4--8 $N$ times. 
\end{enumerate}

We experimented with variants of this procedure, to include weighting the proposed $\theta$ rather than crude accept/reject, and with a variety of feature vectors, distance measures and acceptance thresholds.
We concentrated on $\theta=(\lambda_0, \lambda_1,\mu)$ with the step length and direction change parameters considered to be nuisance. We found the following features to be particularly informative for $\theta$.
\begin{enumerate}
\item The total number of points overall.
\item The number of points not in filaments.
\item The number of filaments.
\item The number of blunt triads amongst points not deemed to be in filaments.
\end{enumerate}
As distance measure we found
\[  d(y_{{\tt sim}},y_{{\tt obs}}) = \left\| \frac{ s_{{\tt sim}}-s_{{\tt obs}}}{s_{{\tt obs}}}\right\|_1 \]
to be effective. An acceptance threshold of 0.5 worked well.

\subsection{Simulations III}

Table \ref{tab:abcsims} summarises some of our simulation results.  Each block is based on 100 simulations, with each simulation using $N=5000$ ABC iterations. The simulated data were generated as in Section \ref{sec:sims1}, with points in a $150 \times 360$ rectangle and $d_0=10$, $\varepsilon=15\pi/180$.

The results in the table were obtained with $\theta$  generated independently at each iteration.  We took $\lambda_0 \sim U(10,110)$, $\lambda_1 \sim U(100,700)$ and $\log \mu \sim U(\log 0.5, \log 5)$.  The procedure is slow, taking about a week to complete all simulations when running on a powerful machine, but without any attempt to parallelise. We obtained similar results when generating $\theta$ as a Markov chain as in Algorithm 2 of \cite{fearnhead12}.  In both cases the results are good and the ABC procedure seems to work well.

\begin{table}
\caption{\label{tab:abcsims} ABC simulation results. SD is standard deviation and RMSE is root mean squared error.}
 \centering
\begin{tabular}{lrrrr} 
  & True & Mean & SD & RMSE \\
$\lambda_0$ & 60 & 61.9 & 18.5 & 21.8 \\
$\mu$ & 2 & 1.97 & 0.59 & 0.78 \\
$\lambda_1$ & 350 & 349.3 & 56.6 & 65.8 \\
\\
$\lambda_0$ & 60 & 61.6 & 16.0 & 18.6 \\
$\mu$ & 3 & 2.87 & 0.60 & 0.73 \\
$\lambda_1$ & 290 & 294.2 & 48.7 & 56.1 \\
\\
$\lambda_0$ & 80 & 79.2 & 18.5 & 20.0 \\
$\mu$ & 2 & 1.94 & 0.52 & 0.61 \\
$\lambda_1$ & 250 & 258.5 & 51.8 & 58.8 \\
\\
$\lambda_0$ & 40 & 43.1 & 18.6 & 21.1\\
$\mu$ & 2 & 1.97 & 0.77 & 0.97 \\
$\lambda_1$ & 450 & 454.0 & 61.4 & 70.6 \\
\end{tabular}
\end{table}

\section{Applications}
\label{sec:apps}

\subsection{Precipitation data}
\label{sec:precipapp}
The first application of our method is on data taken from publicly available climate simulation results made available through the Large Ensemble Community Earth System Model project \citep{kay2015}. 
There are 40 different replications  over a global spatial  grid  of $288 \times 192$ locations.  We concentrate on the annual mean for 2020 and, as suggested by \citet{cast16},  we consider the region between latitudes $-62^\circ$ to $72^\circ$ so as avoid instabilities near the poles. Figure \ref{fig:precip} shows the locations of local minima in the residuals of the first replication, after point-wise standardisation by the mean and standard deviation across replications, which removed latitude, longitude and other fixed effects.  As previously mentioned, such residuals are usually assumed to be realisations of Gaussian random fields \citep{cast16, edwards19}. In the following we used the great circle distance between locations and properly took  into account cyclic behaviour for data on a sphere.

\begin{figure}[h]
	\centering{\includegraphics[width = 0.7\textwidth]{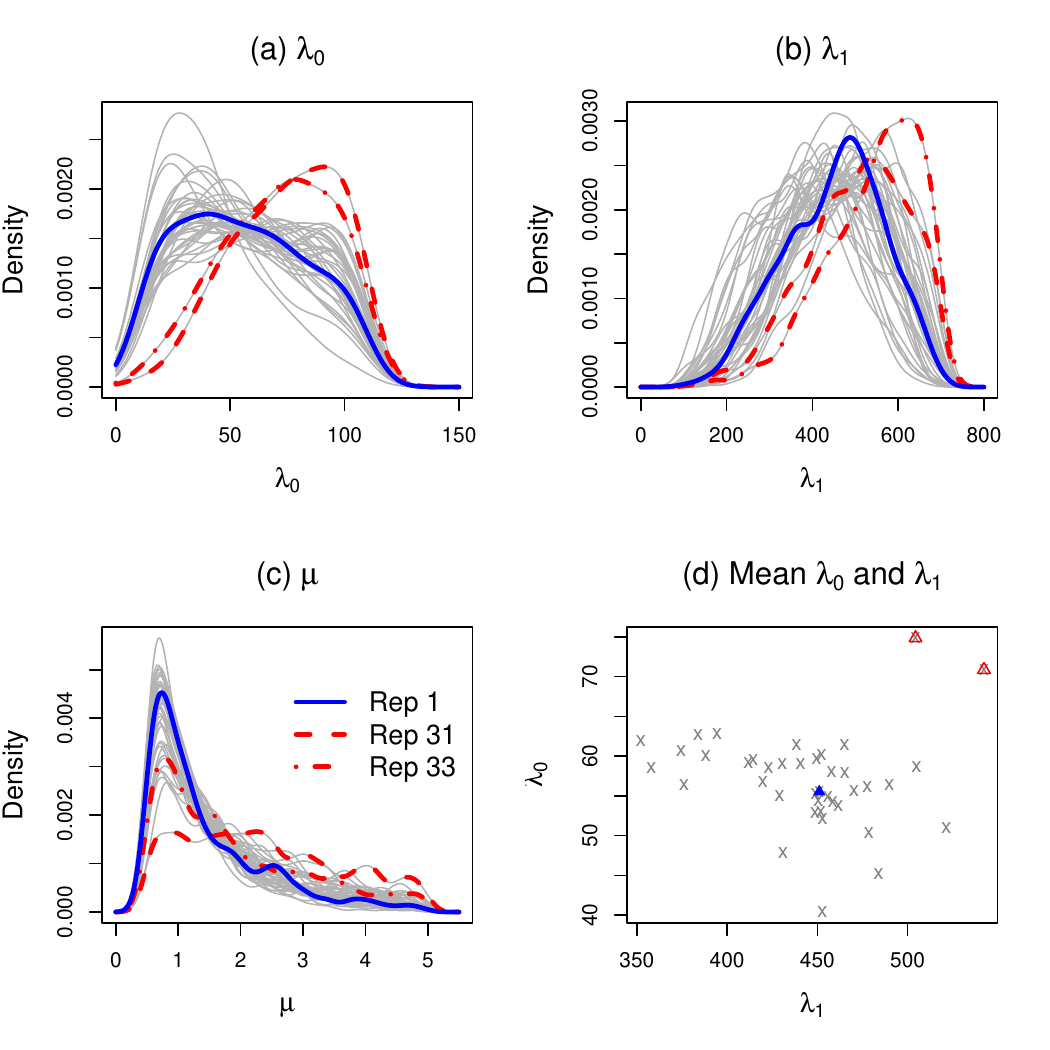}}
	\caption{Posterior densities and mean parameters $\lambda_0$ and $\lambda_1$ for climate data. The bold solid lines in panels (a)--(c), and the filled triangle in panel (d), correspond to the first replication, as in Figure \ref{fig:precip}. The bold broken lines and empty triangles correspond to replications 31 and 33.}
	\label{fig:climres}
\end{figure}

Figure \ref{fig:climres} shows, for all 40 replications, the posterior distributions for the Poisson filament process parameters $\lambda_0$ (number of filaments), $\lambda_1$ (random points) and $\mu$ (filament size parameter). In all replications we have strong evidence that the points do not form a Poisson process and the Poisson filament process is more tenable, pointing to a non-Gaussian residual structure. The median number of   
filaments is 57, the median filament length (remembering the parametrisation as $3+\mu$) is 4.5 and the median number of random points is 450.  Replication 1 is typical of the majority of replications.  It is interesting  to note that two replications stand out as having unusual posterior densities for $\lambda_0$, and to a lesser extent $\lambda_1$.  These are replications 31 and 33.  \cite{baker16} pointed out, after the majority of the data were released, that  these two replications were generated in a different manner to the others.  The authors had previously made public the fact that two of the replications were unusual, without revealing which two these were, and had challenged climate scientists to identify the two unusual ensemble members. They subsequently concluded that skilled researchers could detect the unusual replications, but only after careful and detailed analysis of a large number of variables. 

\subsection{Cold clump data}
\label{sec:coldapp}

\begin{figure}[h]
	\centering{\includegraphics[width = 0.7\textwidth]{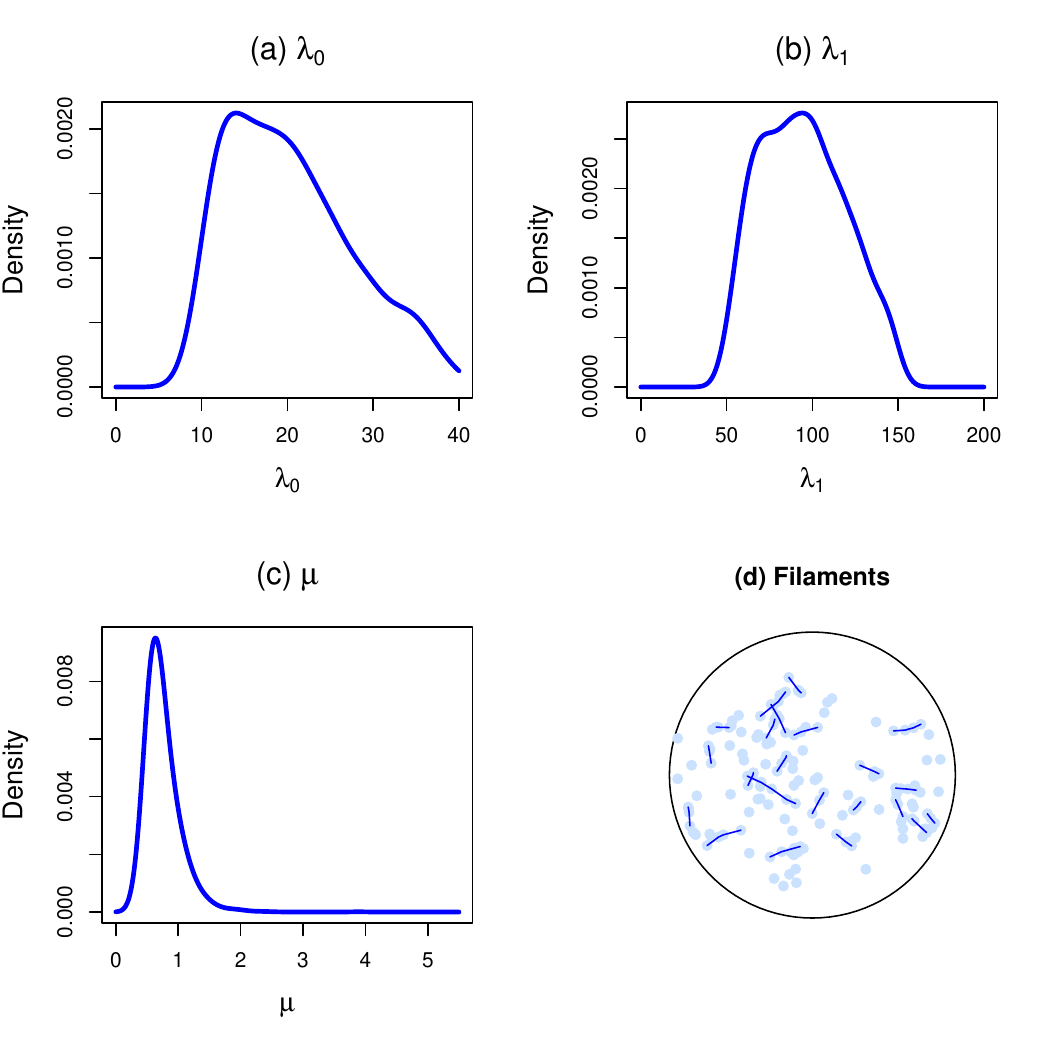}}
	\caption{Posterior densities and estimated filaments for cold clump data.}
	\label{fig:ccres}
\end{figure}

Figure \ref{fig:ccres} shows the posterior densities for $\lambda_0$, $\lambda_1$ and $\mu$ for the cold clump data, together with, in panel (d), the filaments from which the observed feature vector was formed. The posterior median number of   
filaments is 19, the posterior median filament length  is 3.7 and the posterior median number of random points is 93.
The vertical scales in panels (a) and (b) of  Figure \ref{fig:ccres} differ from those in Figure \ref{fig:climres} given the lower numbers of points, but both figures have the same scale in panel (c) as the size parameter $\mu$ has the same interpretation in each application.  We see that for the cold clump data the ratio of noise points to number of filaments is similar to that for the climate data, but the latter has more support for longer filaments. 
With occasional exceptions, the filaments in panel (d) are short, and there is no clear spatial pattern to their aligments. Our evidence points to a mixture of mechanisms to star formation, with some filamentarity but many cold clumps not in evident filaments.

\section{Discussion}

We have considered filamentarity in point process data arising in two quite different applications. In both cases simple diagnostic methods suggest non-random patterns. For the cold clump data we have just a single realisation and we are interested in the specific observed pattern. We conclude that there is support for some filamentarity in star formation within the Milky Way, though many cold clumps, and presumably later stars, are not in observed filaments.  For the climate data  we are interested in characterising the stochastic behaviour of local minima rather than the observed pattern in any single realisation. For this the Poisson filament process seems a realistic modelling approach, leading to a simple parametric summary, formally valid inference and straightforward comparison of realisations. The method successfully picked out two outlying realisations and provides strong evidence of a non-Gaussian residual field. The ABC estimation procedure works well, albeit slowly at present, at least in our coding.

Two extensions might be given consideration.  The first is to consider filamentarity in three dimensions rather than two.  Galactic data are expected to have accurate distance information in the near future, which will mean new cosmic web methodology  will need to be developed.  How best to define filaments in three dimensions needs attention, as will computational issues.

A second extension is to develop a spatio-temporal approach.  We have considered the Large Ensemble  climate data for just one year, 2020,  but annual data are available for model runs up to the year 2100. It would be interesting to investigate how filaments evolve 
and change over time.  

\section*{Appendix A:  $N(\varepsilon, d_0)$}

\begin{figure}[t]
\centering{\includegraphics[width = 0.7\textwidth]{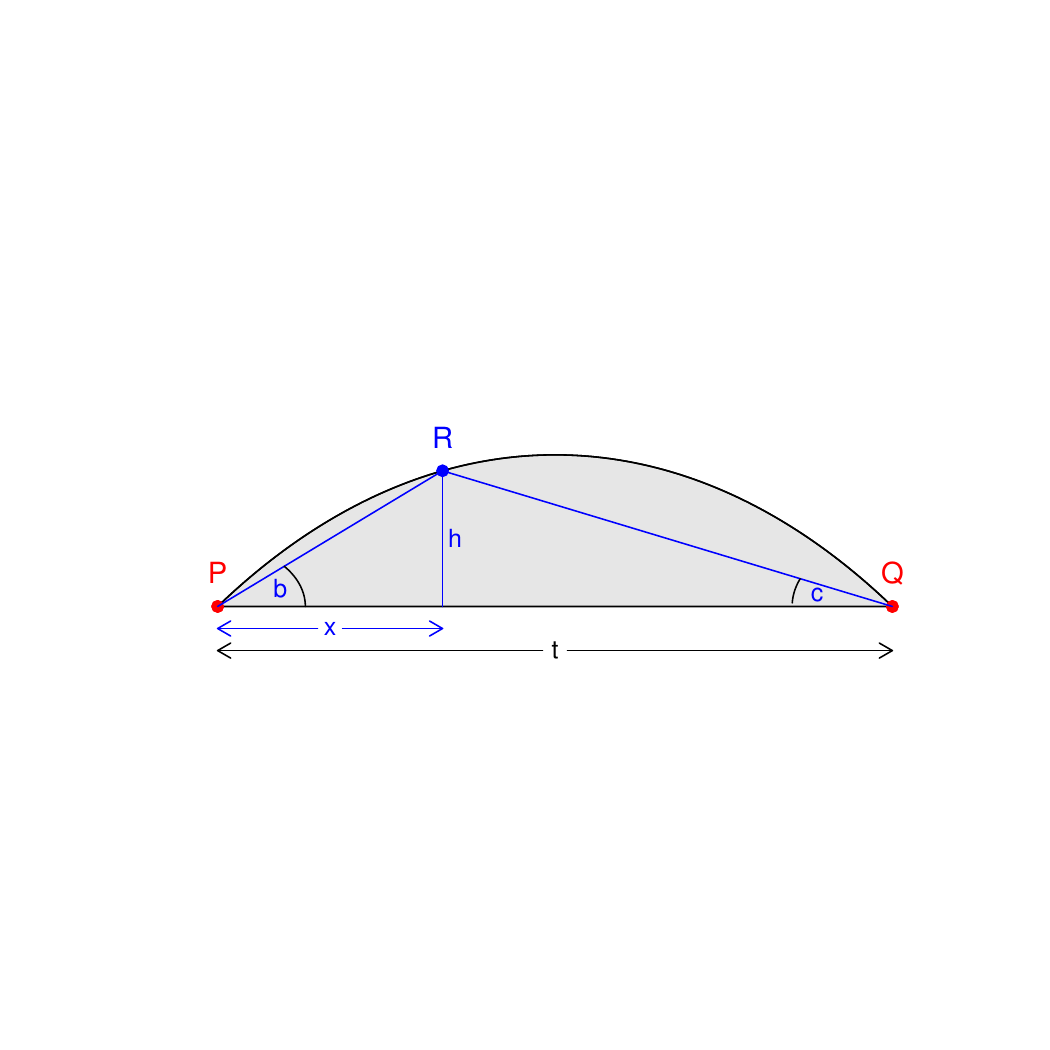}}
  \caption{Central region}
\label{fig:lens}
\end{figure}

We first adapt the methods of \cite{broadbent80} and \cite{kendall80} to find the probability that three points uniformly distributed in an $s \times 1$ rectangle form an $(\varepsilon, d_0)$-blunt triad.  The mean and variance of the number  $N(\varepsilon, d_0)$
of $(\varepsilon, d_0)$-blunt triads formed when $n$ points are uniformly distributed in the rectangle then follows directly from the results of 
\cite{kendall80}. The derivations are given here for completeness.

\subsection*{No length restrictions}

First consider $P$ and $Q$ in Figure \ref{fig:secants} to be fixed. The length between $P$ and $Q$ is $t=t_{PQ}$ and the secant lengths are $u=u_{PQ}$ and $v=v_{PQ}$.  Assume that a third point will form an $\varepsilon$-blunt triad with $P$ and $Q$ if it falls in the shaded region.  To begin with we will not consider edge lengths.

The area of the left wedge is twice the area of the isoceles triangle defined by the dotted line: the excess area of the upper part of the wedge is cancelled by the shortfall in area of the lower part.  Following Broadbent, we can ignore corner effects as these will be $O(\varepsilon^2)$ or smaller for small $\epsilon$. Then using $\tan(\varepsilon)=\varepsilon + O(\varepsilon^3) = \varepsilon + o(\varepsilon)$ we see that the area of the left wedge
is
\[ A_u= 2 \times \left(\frac{1}{2}u^2 \tan(\varepsilon)\right) = u^2 \varepsilon + o(\varepsilon). \]Similarly the area of the right wedge is 
\[ A_v=   v^2 \varepsilon + o(\varepsilon). \]

To obtain the area of the lens region consider the point $R$ in Figure \ref{fig:lens}. This is on the boundary and so the angle of $PQR$ at $R$ is $\pi-\varepsilon$. In turn 
\begin{eqnarray*}
\varepsilon & = & b + c \\
 & = & \tan^{-1}(h/x) + \tan^{-1}(h/(t-x)).
\end{eqnarray*}
The arc tan function also has the series expansion $\tan^{-1}(h/x)= h/x+O((h/x)^3)$ for small $h/x$ and so to first order

\[ \varepsilon  \simeq  \frac{h}{x}+\frac{h}{t-x}
  =  \frac{ht}{x(t-x)}.\]

The area of the lens is twice the area of the upper shaded region.  Consequently, remembering $h=h(x)$ and still to first order:
\begin{eqnarray*}
A_t & = & 2\int_0^t h dx\\
& \simeq & \frac{2\varepsilon}{t} \int_0^t x(t-x) dx \\
 & = & \frac{1}{3}t^2 \varepsilon.
\end{eqnarray*}

The total area of the shaded region is thus $\varepsilon( u^2 + \frac{1}{3}t^2 +v^2)+o(\varepsilon)$
as given in equation (13) of \cite{kendall80}.

\subsection*{Length restrictions included}

Now we impose the edge length restriction: edges adjacent to the largest angle are required to be no longer than $d_0$.  We need to consider separately the cases $t < d_0$, $d_0  \leq t \leq 2d_0$ and $t > 2d_0$.

If $t < d_0$ then all triangles defined over the lens region automatically satisfy the restriction.  Defining $u_{d_0}= \mathrm{min}(u,d_0)$ and $v_{d_0} = \mathrm{min}(v,d_0)$ the relevant total area is $\varepsilon( u_{d_0}^2 + \frac{1}{3}t^2 +v_{d_0}^2)+o(\varepsilon)$.  If $t > 2d_0$ then no triangles  satisfy the restriction.

This leaves $d_0 \leq t < 2d_0$. In this case the third point can only be in the lens part of Figure \ref{fig:secants}. Considering the two triangles formed by the height $h$ in  Figure \ref{fig:lens}, we can write
\[ x^2+h^2 \leq d_0^2 \;\;\;\; \mbox{and} \;\;\;\; (t-x)^2+h^2  \leq d_0^2, \]
and we remember that $h=h(x,t,\varepsilon)$.  Hence the triangle PQR will be 
$(\varepsilon, d_0)$-blunt if and only if $R$ lies in a subset of the lens of the form
\[   t-d_0 + \delta_1 \leq  x \leq d_0 - \delta_2, \]
where $\delta_1$ and $\delta_2$ are both non-negative functions of $t$, $\varepsilon$ and $d_0$.  We know that $h$ can be written as 
\[ h \simeq \frac{x(t-x)}{t}\varepsilon  \]
which has a maximum, at $x=t/2$, of
\[ h_{max} \simeq \frac{t}{4}\varepsilon.  \]
Consequently  the maximum values of both $\delta_1$ and  $\delta_2$ are each  $O(\varepsilon)$. Their minimum values are zero, at $t=d_0$.

Now define 
\begin{eqnarray*}
	A_{t,d_0} & = & 2\int_{ t-d_0 + \delta_1}^{d_0 - \delta_2}   h dx\\
	& = & 2\int_{ t-d_0}^{d_0}   h dx -  2\int_{ t-d_0}^{t-d_0+\delta_1}   h dx-  2\int_{d_0-\delta_2}^{d_0}   h dx.
\end{eqnarray*}
Consider the second integral.  Noting that the integrand is non-negative
\[	\int_{ t-d_0}^{t-d_0+\delta_1} h dx  \leq    \int_{ t-d_0}^{t-d_0+\delta_1} h_{max} dx  =  h_{max} \delta_1 = O(\varepsilon^2). \]
The same property holds for the third integral.  These terms are thus small in comparison with the first integral, which is $O(\varepsilon)$.  Thus to first order in $\varepsilon$
\begin{eqnarray*}
 A_{t,d_0}& = & 2 \int_{ t-d_0}^{d_0} h dx  
  =  2 \varepsilon\int_{ t-d_0}^{d_0}  \dfrac{x(t-x)}{t} dx \\
 & = & \left(2d_0^2- \dfrac{t^2}{3} - \dfrac{4d_0^3}{3t}\right)\varepsilon.
\end{eqnarray*}

Hence, conditional on P and Q,  the triangle PQR will be $(\varepsilon, d_0)$-blunt if R falls in a region with area approximated by $H(P,Q)\varepsilon$, where

\[ H(P,Q) =  \left\{ \begin{array}{ll}
u_{d_0}^2 + \frac{1}{3}t^2 +v_{d_0}^2  & \mbox{if } t \leq d_0,\\
\\
2d_0^2- \dfrac{t^2}{3} - \dfrac{4d_0^3}{3t} &  \mbox{if } d_0 < t \leq 2d_0,\\
\\
0 &  \mbox{if }  t \geq 2d_0.\\
\end{array} \right.
\]

\subsection*{Mean and variance of $N(\varepsilon, d_0)$}

Now consider $n$ points independently and uniformly distributed over region $K$. Label the triads formed from the points as $1,2,\ldots$  and define
\[ I_j  = \left\{ \begin{array}{ll}
1 & \mbox{if triad } j \mbox{ is }(\varepsilon, d_0)\mbox{-blunt},\\
0 & \mbox{otherwise}. \end{array}\right.\]

Clearly 
\[ \mathbb{E}(I_j)=\alpha \varepsilon +o(\varepsilon), \hspace*{1cm} \mbox{Var}(I_j)=
\alpha \varepsilon(1-\alpha \varepsilon )+o(\epsilon), \]
where $\alpha=\alpha(K)$ is defined at \eqref{eqn:lambda}.

Consider pairs of triads $j$ and $k$ as being in one of three groups:
\begin{itemize}
\item[] $\G_0$:  no points in common,
\item[] $\G_1$:  one point in common,
\item[] $\G_2$:  two points in common.
\end{itemize} 
Given that $N(\varepsilon, d_0)= \sum_{j} I_j$ we have
\begin{eqnarray*}
\mbox{Var}(N(\varepsilon, d_0)) & = & \sum_j \mbox{Var}(I_j) + \sum_{j,k} \mbox{Cov}(I_j,I_k)\\
 & = &  \mathbb{E}[N(\varepsilon, d_0)](1-\alpha \varepsilon ) + \sum_{j,k \in \G_0} \mbox{Cov}(I_j,I_k)\\
 &  & 
+\sum_{j,k \in \G_1}\mbox{Cov}(I_j,I_k) +\sum_{j,k \in G_2} \mbox{Cov}(I_j,I_k) + o(\varepsilon^2) \\
 & = &  \mathbb{E}[N(\varepsilon, d_0)](1-\alpha \varepsilon ) + \mid \G_0 \mid (\eta-\alpha^2)
\varepsilon^2+ \mid \G_1 \mid (\beta-\alpha^2)\varepsilon^2\\
 & & + \mid \G_2 \mid (\gamma-\alpha^2)\varepsilon^2+o(\varepsilon^2),
\end{eqnarray*}
where
\begin{eqnarray*}
\Pr(\mbox{two triads with no shared point are both } (\varepsilon, d_0)\mbox{-blunt}) & = &
\eta\varepsilon^2 + o(\varepsilon^2)\\
\Pr(\mbox{two triads with one shared point are both } (\varepsilon, d_0)\mbox{-blunt}) & = & \beta \varepsilon^2 + o(\varepsilon^2)\\
\Pr(\mbox{two triads with two shared points are both } (\varepsilon, d_0)\mbox{-blunt}) & = &\gamma \varepsilon^2+ o(\varepsilon^2).\\
\end{eqnarray*}

As points are independent we have $\eta=\alpha^2$ and there is zero covariance between $I_j$ and $I_k$  for $j,k \in \G_0$.

For $\G_1$, label the triads $PQ_1R_1$ and $PQ_2R_2$, where $P$ is the shared point.  Conditional upon that point, the probability that $PQ_1R_1$ is $(\varepsilon, d_0)$-blunt  (to within $o(\varepsilon)$) is
$\varepsilon\mathbb{E}H(P,Q_1)/|K|$ and the probability that $PQ_2R_2$ is $\varepsilon$-blunt is 
$\varepsilon\mathbb{E}H(P,Q_2)/|K|$.  These expectations are the same and hence
\[ \beta = \beta(K) =  \frac{1}{|K|^2}\mathbb{E}_P\{[\mathbb{E}_Q H(P,Q)]^2\}.\]

Turning to $\G_2$, label the triads $PQR_1$ and $PQR_2$.  Given $P$ and $Q$, each triad is 
 $(\varepsilon, d_0)$-blunt with probability  $\varepsilon H(P,Q)/|K|$ (to within $o(\varepsilon)$), independently of each other.  Thus
\[ \gamma = \gamma(K) =  \frac{1}{|K|^2}\mathbb{E}_P\mathbb{E}_Q\{ H(P,Q)^2\}.\]

Finally we need to consider $\mid \G_1 \mid$ and $\mid \G_2 \mid$.  For these we pick one triad at random and then pick the additional points of the second triad from the $n-3$ remaining points, leading to

\[ \mid \G_1 \mid = 3\begin{pmatrix}
n\\
3
\end{pmatrix}\begin{pmatrix}
n-3\\
2
\end{pmatrix}
\hspace*{1cm}\mid \G_2 \mid = 3\begin{pmatrix}
n\\
3
\end{pmatrix}\begin{pmatrix}
n-3\\
1
\end{pmatrix}.
\]
The multipliers 3 arise because there are three choices  for the unshared point in the first triad in $\G_1$, and three for the shared point in $\G_2$.

Putting the above together leads to \eqref{eqn:varN}, which is the edge-restricted version of  Kendall \& Kendall's expression (6) for Var$(N(\varepsilon))$.

\section*{Funding information}

There is no funding to report.

\section*{Data availability statement}

The data used in the paper are available from the authors on request.

\bibliographystyle{agsm}
\bibliography{filrefs}

\end{document}